\documentclass{aastex}
\usepackage{spr-astr-addons}
\usepackage{url}

\RequirePackage{color}

\newcommand{\emaila}{mohsen.shadmehri@dcu.ie}

\begin{document}

\title{Hot accretion with outflow and thermal conduction}
\shorttitle{Hot accretion with outflow and thermal conduction}
\shortauthors{M. Shadmehri}

\author{Mohsen Shadmehri\altaffilmark{1,2}}

\email{\emaila}

\altaffiltext{1}{School of Mathematical Sciences, Dublin City University, Glasnevin, Dublin 9, Ireland}
\altaffiltext{2}{Department of Physics, School of Science, Ferdowsi University, Mashhad, Iran}


\begin{abstract}
We present self-similar solutions for advection -dominated accretion flows with thermal conduction in the presence of outflows. Possible effects of outflows on the accretion flow are parametrized and a saturated form of thermal conduction, as is appropriate for the weakly-collisional regime of interest, is included in our model. While the cooling effect of outflows is noticeable, thermal conduction provides an extra heating source. In comparison to accretion flows without winds, we show that the disc rotates faster and becomes cooler  because of the angular momentum and energy flux which are taking away by the winds. But thermal conduction opposes the effects of winds and not only decreases the rotational velocity, but increases the temperature. However, reduction of the surface density and the enhanced  accretion velocity are amplified by both of the winds and the thermal conduction. We find  that for stronger outflows, a higher level of saturated thermal conduction is needed to significantly modify the physical profiles of the accretion flow.
\end{abstract}

\keywords{accretion - accretion discs - black hole physics}


\section{Introduction}
The thin accretion disk model describes flows in which the viscous heating of the gas radiates out of the system immediately after generation (Shakura \& Sunyaev 1973). However, another kind of accretion has been studied during recent years where radiative energy losses are small so that most of the energy is advected with the gas. These Adection-Dominated Accretion Flows (ADAF) occur in two regimes depending on the mass accretion rate and the
optical depth. At very high mass accretion rate, the optical depth becomes very high and the radiation can be trapped in the gas. This type of accretion which is known under the name 'slim accretion disk' has been studied in detail by Abramowicz et al. (1988). But when the accretion rate is very small and the optical depth is very low, we may have another type of accretion (Narayan \& Yi 1994; Abramowitz et al. 1995; Chen 1995). However, numerical simulations of radiatively inefficient accretion flows revealed that low viscosity flows are convectively unstable and convection strongly influences the global properties of the flow (e.g., Igumenshchev, Abramowicz, \& Narayan 2000). Thus, another type of accretion flows has been proposed, in which convection plays as a dominant mechanism of transporting angular momentum and the local released viscous energy (e.g., Narayan, Igumenshchev, \& Abramowicz 2000).

This diversity of models tells us that modeling the hot accretion flows is a challenging and controversial problem. We think, one of the largely neglected physical ingredient in this field, is {\it thermal conduction}. But a few authors tried to study the role of "turbulent" heat transport in ADAF-like flows (Honma 1996; Manmoto et al. 2000). Since thermal conduction acts to oppose the formation of the temperature gradient that causes it, one might expect that the temperature and density profiles for accretion flows in which thermal conduction plays a significant role to appear different compared to those flows in which thermal conduction is less effective. Just recently, Johnson \& Quataert (2007) studied the effects of electron thermal conduction on the properties of hot accretion flows, under the assumption of spherical symmetry. In another interesting analysis, Tanaka \& Menou (2006) showed that thermal conduction affects  the global properties of hot accretion flows substantially. They generalized standard ADAF solutions to include a saturated form of thermal conduction. In the second part of their paper, a set of two dimensional self-similar solutions of ADAFs in the presence of thermal conduction has been presented. The role of conduction is in providing the extra degree of freedom necessary to launch thermal outflows according to their 2D solutions.

On the other hand, ADAFs with winds or outflows have been studied extensively during recent years, irrespective of possible driven mechanisms of winds. But thermal conduction has been neglected in all these ADAFs solutions with winds. In advection-dominated inflow-outflow solutions (ADIOS), it is generally assumed that the mass flow rate has a power-law dependence on radius, with the power law index, $s$, treated as a parameter (e.g., Blandford \& Begelman 1999; Quataert \& Narayan 1999; Beckert 2000; Misra \& Taam 2001; Fukue 2004; Xie \& Yuan 2008). Beckert (2000) presented self-similar solutions for ADAFs with radial viscous force in the presence of outflows from the accretion flow or infall. Turolla \& Dullemond (2000) investigated how, and to what extent, the inclusion of the source of ADAF material affects the Bernoulli number and the onset of a wind. In their model, the accretion rate decreases with radius ($s<0$). Misra \& Taam (2001) studied the effect of a possible hydrodynamical wind on the nature of hot accretion disc solutions. They showed that their solutions are locally unstable to a new type of instability called wind-driven instability, in which the presence of a wind causes the disc to be unstable to long-wavelength perturbations of the surface density. Kitabatake, Fukue \& Matsumoto (2002) studied supercritical accretion disc with winds, though angular momentum loss of the disc, because of the winds, has been neglected. Comparisons with observations reveal that the X-ray spectra of such a wind-driven self-similar flow,  can explain the observed spectra of black hole candidates in quiescence (Quataert \& Narayan 1999; Yuan, Markoff \&  Falcke 2002). Lin, Misra \& Taam (2001) find that the spectral characteristics of high-luminosity black hole systems suggest that winds may be important for these systems as well.

Considering extensive works on hot accretion flows with winds and significant role of thermal conduction in deriving outflows (Tanaka \& Menou 2006), we study ADAFs with outflows {\it and} thermal conduction using height-integrated set of equations. A phenomenological way is adopted in which we parameterize the rate at which mass, angular momentum and energy are extracted by outflow or wind. In the next section, we present basic equations of the model. Self-similar solutions are investigated in section 3. The paper concludes with a summary of the results in section 4.

\section{General formulation}
We consider an accretion disc that is axisymmetric and geometrically thin, i.e. $H/R < 1$. Here $R$ and $H$ are, respectively, the disk radius and the half-thickness. The disc is supposed to be turbulent and possesses an effective turbulent viscosity. Consider stationary height-integrated equations described an accretion flow onto a central object of mass $M_{\ast}$. The continuity equations reads

\begin{equation}
\frac{\partial}{\partial R} (R\Sigma v_{\rm R}) + \frac{1}{2\pi} \frac{\partial \dot{M}_{\rm w}}{\partial R} = 0,\label{eq:con}
\end{equation}
where $v_{\rm R}$ is the accretion velocity ($v_{\rm R}<0$) and $\Sigma = 2\rho H$ is the surface density at a cylindrical radius $R$. Also, $\rho$ is the midplane density of the disc and the mass loss rate by outflow/wind is represented by $\dot{M}_{\rm w}$. So,
\begin{equation}
\dot{M}_{\rm w}(R) = \int 4\pi R' \dot{m}_{\rm w} (R') dR',\label{eq:mdot}
\end{equation}
where $\dot{m}_{\rm w} (R)$ is mass loss rate per unit area from each disc face.

Xie \& Yuan (2008) derived the height-integrated accretion equations including the coupling between the inflow and outflow, to investigate the influence of outflow on the dynamics of hot inflow. They showed that under reasonable assumptions to the properties of outflow, the main influence of of outflow can be properly included by adopting a radius dependent mass accretion rate. We write the dependence of the accretion rate $\dot{M}$ as follows (e.g., Blandford \& Begelman 1999)
\begin{equation}
\dot{M}=-2\pi R \Sigma v_{\rm R} = \dot{M}_{0}(\frac{R}{R_{0}})^{s},\label{MMdot}
\end{equation}
where $\dot{M}_{0}$ is the mass accretion rate at the outer boundary $R_{0}$. The parameter $s$ describes how the density profile and the accretion rate are modified. In this paper, typical values of $s$ considered are between $s=0$ (no winds) and $s=0.3$ (moderately strong wind). The above prescription for mass accretion rate has been used widely for $s>0$ (e.g.,  Quataert \& Narayan 1999; Beckert 2000; Misra \& Taam 2001; Fukue 2004) or $s<0$ (Turolla \& Dullemond 2000). Considering equations (\ref{eq:con}), (\ref{eq:mdot}) and (\ref{MMdot}), we can obtain
\begin{equation}
\dot{m}_{\rm w}=\frac{s}{4\pi R_{0}^{2}} \dot{M}_{0} (\frac{R}{R_{0}})^{s-2}.
\end{equation}

The equation of motion in the radial direction is
\begin{equation}
v_{\rm R}\frac{dv_{\rm R}}{dR}=R(\Omega^{2}-\Omega_{\rm K}^{2})-\frac{1}{\rho}\frac{d}{dR}(\rho c_{\rm s}^{2}),
\end{equation}
where $\Omega$ is angular velocity and $\Omega_{\rm K}=\sqrt{GM_{\ast}/R^{3}}$ represents Keplerian velocity. Also, $c_{\rm s}$ is sound speed and from vertical hydrostatic equilibrium, we have $H=c_{\rm s}/\Omega_{\rm K}$.

Similarly, integration over $z$ of the azimuthal equation of motion gives (e.g., Knigge 1999)
\begin{equation}
R\Sigma v_{\rm R} \frac{d}{dR} (R^{2}\Omega) = \frac{d}{dR}(R^{3}\nu \Sigma \frac{d\Omega}{dR})-\frac{(lR)^{2}\Omega}{2\pi}\frac{d\dot{M}_{\rm w}}{dR},
\end{equation}
where the last term of right hand side represents angular momentum carried by the outflowing material. Here, $l=0$ corresponds to a non-rotating wind and $l=1$ to outflowing material that carries away the specific angular momentum it had at the point of ejection (Knigge 1999). Also, $\nu$ is a kinematic viscosity coefficient and we assume
\begin{equation}
\nu = \alpha c_{\rm s} H,
\end{equation}
where $\alpha$ is a constant less than unity (Shakura \& Sunyaev 1973).

In order to implement thermal conductivity correctly it is essential to know whether the mean free path is less
than (or comparable to) the scale length of the temperature gradient. For electron mean free path which are greater than the scale length of the temperature gradient the thermal conductivity is said to 'saturate' and the heat flux approaches a limiting value (Cowie \& McKee 1977). But when the mean free paths are much less than the temperature gradient the heat flux depends on the coefficient of thermal conductivity and the temperature gradient. Generally, thermal conduction transfers heat so as to oppose the temperature gradient which causes the transfer. Tanaka \& Menou  (2006) discussed hot accretion likely proceed under weakly-collisional conditions in these systems. Thus, a saturated form of "microscopic" thermal conduction is physically well-motivated, as we apply in this study. However, one of the primary problems for studying the effects of thermal conduction in plasmas is the unknown value of the thermal conductivity.

Now, we can write the energy equation considering energy balance in the system. We assume the generated energy due to viscous dissipation and the heat conducted into the volume is concerned are balanced by the advection cooling and energy loss of outflow. Thus,

\begin{displaymath}
\frac{\Sigma v_{\rm R}}{\gamma -1}\frac{dc_{\rm s}^{2}}{dR}-2Hv_{\rm R}c_{\rm s}^{2}\frac{d\rho}{dR}=f\frac{\alpha\Sigma c_{\rm s}^{2}R^{2}}{\Omega_{\rm K}}(\frac{d\Omega}{dR})^{2}
\end{displaymath}
\begin{equation}
-\frac{2H}{R}\frac{d}{dR}(RF_{\rm s})-\frac{1}{2}\eta \dot{m}_{\rm w}(R) v_{\rm K}^{2}(R),
\end{equation}
where the second term on right hand side represents energy transfer due to the thermal conduction and $F_{\rm s} = 5 \phi_{\rm s} \rho c_{\rm s}^{3}$ is the saturated conduction flux on the direction of the temperature gradient (Cowie \& McKee 1977). Dimensionless coefficient $\phi_{\rm s}$ is less than unity. Also, the last term on right hand side of energy equation is the energy loss due to wind or outflow (Knigge 1999). Depending on the energy loss mechanism, dimensionless parameter $\eta$ may change. We consider it as a free parameter of our model so that larger $\eta$ corresponds to more energy extraction from the disc because of the outflows (Knigge 1999).

\section{Self-similar solutions}

A self-similar solution is not able to describe the {\it global} behaviour of an accretion flow, because no boundary condition has been taken into account. However, as long as we are not interested in the behaviour of the flow at the boundaries, such a solution describes correctly the true solution asymptotically at large radii. We assume that each physical quantity can be expressed as a power law of the radial distance, i.e. $R^{\nu}$, where the power index $\nu$ is determined for each physical quantity self-consistently. The solutions are
\begin{equation}
\Sigma (R) = \omega_{0} \Sigma_{0}( \frac{R}{R_{0}})^{s-\frac{1}{2}},
\end{equation}
\begin{equation}
\Omega (R) = \omega_{1} \sqrt{\frac{GM_{\ast}}{R_{0}^{3}}} (\frac{R}{R_{0}})^{-3/2},
\end{equation}
\begin{equation}
v_{\rm R}(R) = -\omega_{2} \sqrt{\frac{GM_{\ast}}{R_{0}}} (\frac{R}{R_{0}})^{-1/2},
\end{equation}
\begin{equation}
P(R) = \omega_{3} P_{0} (\frac{R}{R_{0}})^{s-\frac{3}{2}},
\end{equation}
\begin{equation}
c_{s}^{2}(R) = \frac{\omega_{3}}{\omega_{0}} (\frac{GM_{\ast}}{R_{0}}) (\frac{R}{R_0})^{-1},
\end{equation}
\begin{equation}
H(R) = \omega R_{0} (\frac{R}{R_0}),
\end{equation}
where $\Sigma_0$ and $R_{0}$ provide convenient units with which the equations can be written in the non-dimensional forms. Note that $P(R)$ is the height-integrated pressure. By substituting the above self-similar solutions into the dynamical equations of the system, we obtain the following system of dimensionless equations, to be solved for $\omega$, $\omega_0$, $\omega_1$, $\omega_2$ and $\omega_3$:

\begin{equation}
\omega_{0}\omega_{2}=\dot{m},
\end{equation}
\begin{equation}
\omega_{0}\omega^{2}-\omega_{3}=0,
\end{equation}
\begin{equation}
-\frac{\omega_{2}^{2}}{2}=\omega_{1}^{2}-1+(\frac{5}{2}-s)\frac{\omega_{3}}{\omega_{0}},
\end{equation}
\begin{equation}
\frac{1}{2}\dot{m}-\frac{3}{2}(s+\frac{1}{2})\alpha \omega_{3}-sl^{2}\dot{m}=0,
\end{equation}
\begin{displaymath}
\left [\frac{1}{\gamma -1}+(s-\frac{3}{2}) \right ]\omega_{2}\omega_{3}=\frac{9}{4}\alpha f \omega_{1}^{2}\omega_{3}+5(2-s)\phi_{\rm s}
\end{displaymath}
\begin{equation}
\times\omega_{3}\sqrt{\frac{\omega_3}{\omega_0}}-\frac{\eta s \dot{m}}{4},
\end{equation}
where $\dot{m}=\dot{M}_{0}/(2\pi R_{0}\Sigma_{0}\sqrt{GM_{\ast}/R_{0}})$ is the nondimensional mass accretion rate. After algebraic manipulations, we obtain a forth order algebraic equation for $\omega$:
\begin{displaymath}
 \frac{9\alpha^2}{8} (\displaystyle\frac{1+2s}{1-2sl^2})^{2} \omega^4 +  [ \frac{5}{2} - s + \frac{1+2s}{f(1-2sl^{2})}(\frac{2s}{3}+
\end{displaymath}
\begin{equation}
\displaystyle \frac{5/3 - \gamma}{\gamma -1})  ] \omega^2 + \frac{20 (s-2)\phi_{\rm s}}{9 \alpha f} \omega + \frac{\eta s}{6f}(\frac{1+2s}{1-2sl^2})-1=0,\label{eq:main}
\end{equation}
and the rest of the physical variables are
\begin{figure*}
\vspace*{+100pt}
\includegraphics[scale=0.8]{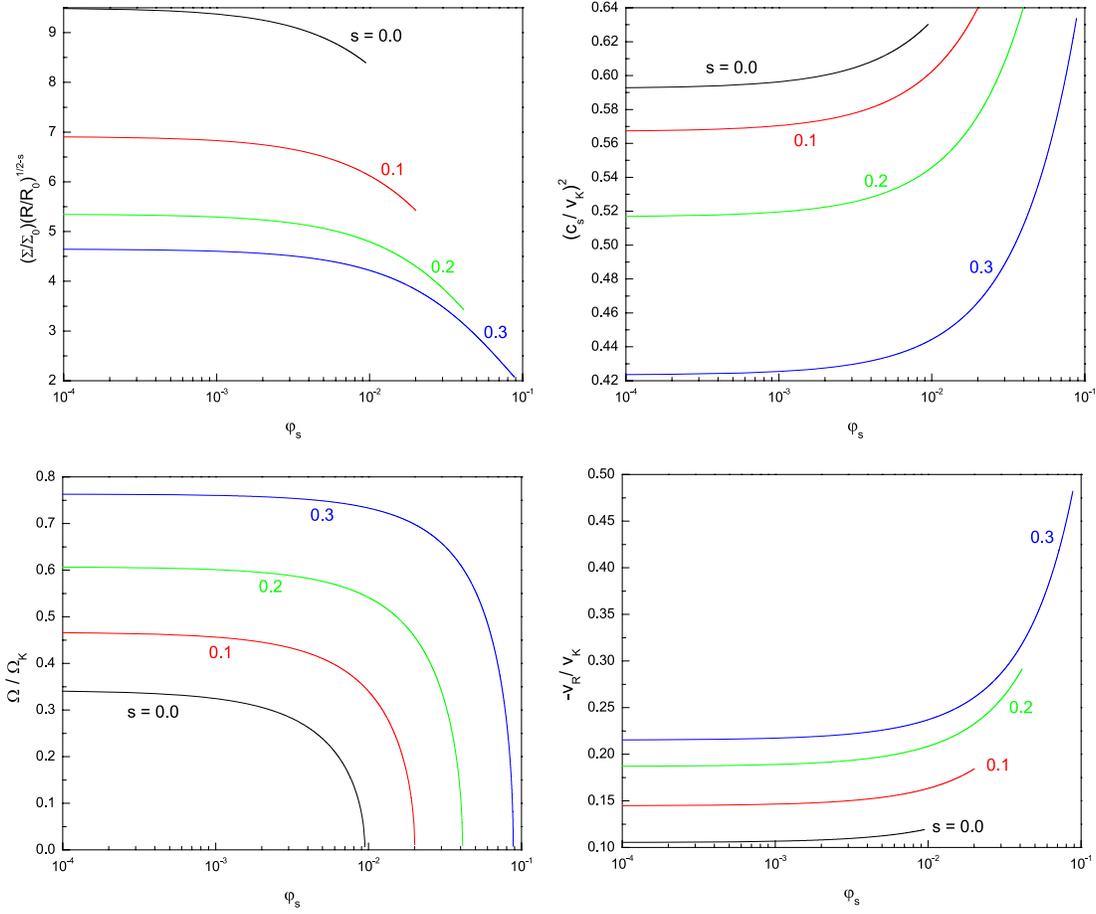}
\caption{Profiles of the physical variables of the accretion disc versus the saturation constant $\phi_{\rm s}$, taking  $\alpha=0.2$, $\gamma=1.5$, $l=1$, $\eta=1$ and $f=1$. Each curve is labeled by its corresponding exponent $s$.} \label{fig:1}
\end{figure*}

\begin{figure*}
\includegraphics[scale=0.8]{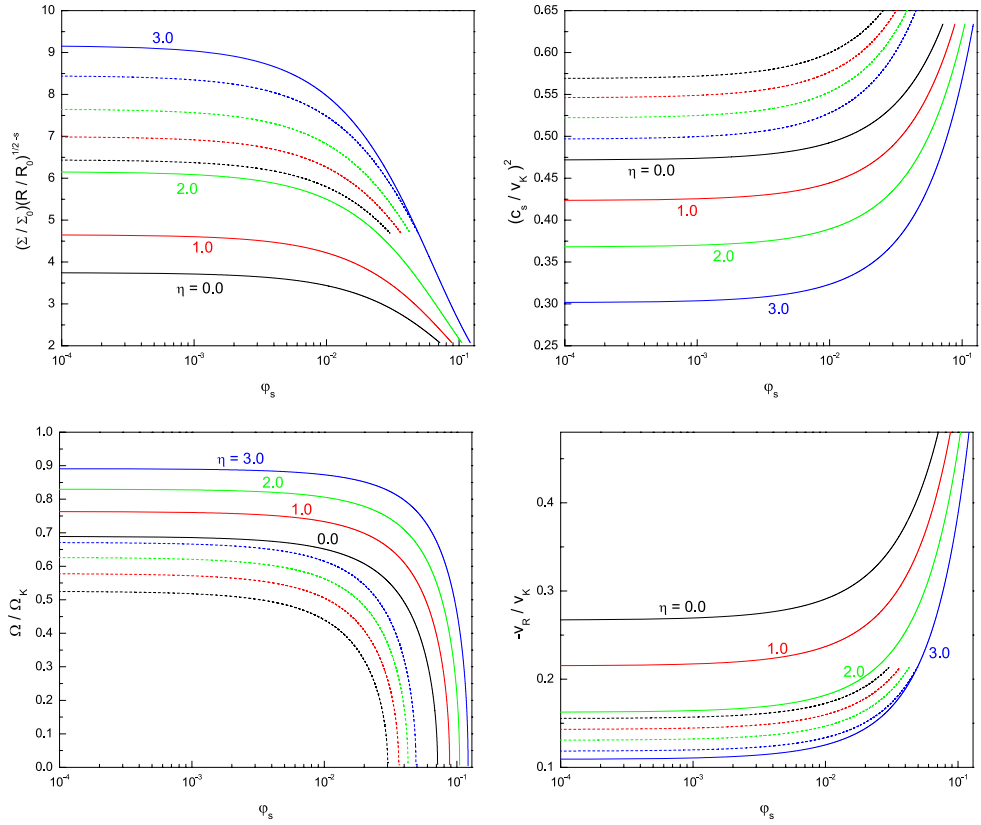}
\caption{The same as Figure \ref{fig:1}, but  $s=0.3$ and each curve is labeled by its corresponding coefficient $\eta$. Solid curves are for $l=1$ and dashed lines represent solutions with $l=0$.} \label{fig:2}
\end{figure*}
\begin{equation}
\omega_{0}=\frac{2}{3\alpha}(\frac{1-2sl^2}{1+2s})\dot{m}\omega^{-2},\label{eq:omega0}
\end{equation}
\begin{equation}
\omega_{1}=\displaystyle\sqrt{1-(\frac{5}{2}-s)\omega^{2}-\frac{9\alpha^{2}}{8}(\displaystyle\frac{1+2s}{1-2sl^{2}})^{2}\omega^{4}},\label{eq:omega1}
\end{equation}
\begin{equation}
\omega_{2}=\frac{3\alpha}{2}(\frac{1+2s}{1-2sl^{2}})\omega^{2},
\end{equation}
\begin{equation}
\omega_{3}=\frac{2}{3\alpha}(\frac{1-2sl^2}{1+2s})\dot{m}.
\end{equation}

We can solve algebraic equation (\ref{eq:main}) numerically and clearly only real roots which correspond to  positive $\omega_{1}^{2}$ are physically acceptable. Without mass outflow and thermal conduction, i.e. $s=l=\eta=0$ and $\phi_{\rm s}=0$, equation (\ref{eq:main}) and similarity solutions reduce to the standard ADAF solutions (Narayan \& Yi 1994). Also, in the absence of wind but with thermal conduction, equation (\ref{eq:main}) reduces to  equation (17) of Tanaka \& Menou (2006). But our main algebraic equation includes both outflows and thermal conduction.

Now we can analysis behavior of the solutions in the presence of the wind and thermal conduction. Our primary goal is to consider the effects of winds and thermal conduction via parameters $s$, $l$, $\eta$ and $\phi_{\rm s}$. First, we can summarize  typical behavior of the standard ADAF solutions   as follows (Narayan \& Yi 1994): (a) The surface density increases with the accretion rate, and decreases with the viscosity coefficient $\alpha$; (b) But the radial velocity is directly proportional to the viscosity coefficient; (c) The gas rotates with sub-Keplerian angular velocity, more or less independent of the coefficient $\alpha$; and finally (d) the opening angle of the disc is fixed, independent of $\alpha$ and $\dot{m}$. Mass outflows and thermal conduction may modify these behaviors according to our solutions. Equation (\ref{eq:omega0}) shows that surface density is directly proportional to the mass accretion rate. But dependence of the surface density on the exponent of accretion rate $s$ is determined by solving equation (\ref{eq:main}). The rotational and the radial velocities are both independent of nondimensional mass accretion rate $\dot{m}$. Also,  dependence of the radial velocity on the viscosity coefficient is determined by the main algebraic equation. In the absence of mass outflows and thermal conduction, the opening angle of the disc is independent of the accretion rate and the viscosity coefficient and this can be understood  from equation (\ref{eq:main})  by setting $s=l=\eta=\phi_{\rm s}=0$. But depending on the angular momentum and energy exchanges due to the wind (i.e. $l$ and $\eta$) and the variable accretion rate (i.e. $s$) and the thermal conduction (i.e. $\phi_{\rm s}$) the thickness of the disc may change according to the acceptable roots of equation (\ref{eq:main}). We can see that the rotational velocity is sub-Keplerian according to equation (\ref{eq:omega1}).

Figure \ref{fig:1} shows profiles of the physical variables versus thermal conduction coefficient $\phi_{\rm s}$ for different accretion rate exponent, i.e. $s=0$, $0.1$, $0.2$ and $0.3$. The value of $s$ measures the strength of outflow, and a larger $s$ denotes a stronger outflow. The other input parameters are $\alpha=0.2$, $\gamma=1.5$, $f=1$ and $l=\eta=1$ and each curve is labeled by its corresponding $s$. Recent work by Sharma et al. (2006) suggest that the viscosity parameter in a hot accretion flow will be larger than in a standard thin disc. If $\alpha$ is much smaller than $0.25$, the maximum accretion rate up to which the ADAF solution is possible, decreases significantly and the maximum luminosity of the models becomes much smaller than the observed luminosities (Quataert \& Narayan 1999). So, $\alpha=0.2$ is probably not unrealistic for our analysis. Clearly profiles with $s=0$ represent no-wind solutions (Tanaka \& Menou 2006). We can see that ADAFs with winds rotate more quickly than those without winds. Also, the viscous dissipation is expected to be larger in the presence of winds and outflows. Strong-wind models have a lower surface density than weak-wind models. While solutions without thermal conduction are recovered at small $\phi_{\rm s}$ values, we can see significant deviations as $\phi_{\rm s}$ increases. Note that for a given set of the input parameters, the solutions reach to a non-rotating limit at a specific value of $\phi_{\rm s}$ which we denote it by $\phi_{\rm s}^{\rm c}$. We can not extend the profiles beyond $\phi_{\rm s}^{\rm c}$, because equation (\ref{eq:omega1}) gives a negative $\omega_{1}^{2}$ which is clearly unphysical. As profiles of Figure \ref{fig:1} show the critical magnitude of conduction viscosity  $\phi_{\rm s}^{\rm c}$ for which the solutions tend to non-rotating limit highly depends on the accretion rate exponent $s$. In fact, higher values of $s$ correspond to larger $\phi_{\rm s}^{\rm c}$.

When there is mass outflow, one can expect lower surface density. Figure \ref{fig:1} shows for non-zero $s$, surface density is lower than the standard ADAF solution and  for stronger outflows, this reduction of the surface density is more evident. However,  profile of the surface density is not affected by $\phi_{\rm s}$ when this parameter is small. But as conduction coefficient $\phi_{\rm s}$ increases, surface density decreases. In particular, this reduction due to the thermal conduction is more significant for larger accretion rate exponent $s$.  In other words, when the disc losses more mass because of the winds or outflows,  surface density  decreases more significantly due to the thermal conduction in comparison  to the no-wind solutions.

Actually, outflows play as a cooling agent and thermal conduction provides extra heating and there is a competition between these physical factors. While  effects of winds and conduction on the profiles of the surface density and the accretion velocity are similar, their effects on the rotational velocity and the temperature oppose each other. Both winds and thermal conduction lead to enhanced accretion velocity and reduced surface density. Although winds increases rotational velocity and decreases the temperature, thermal conduction not only decreases rotational velocity but increases the temperature too.

Angular momentum conservation implies $(1-2sl^2)>0$ which is trivially valid for non-rotating winds (i.e. $l=0$). But for rotating winds, this inequality implies $s<1/2$. Figure \ref{fig:2} show physical profiles for various $\eta$ and $l$, taking $s=0.3$, $\alpha=0.2$, $\gamma=1.5$ and $f=1$. Profiles corresponding to the rotating winds (i.e $l=1$) are shown by solid curves, but dashed curves show non-rotating solutions ($l=0$). Each curve is labeled by its corresponding coefficient $\eta$. Higher this parameter, more energy flux is taking away by winds. Solutions with rotating winds are more sensitive to the variations of the parameter $\eta$ according to the plots of Figure \ref{fig:2}. As more energy flux is extracted from the disc by winds (i.e. higher $\eta$), a lower level of the saturated thermal conduction is enough to significantly modify physical profiles. In particular, this behavior is more evident for solutions with rotating winds.

\section{Discussion and Summary}

Theoretical arguments and observations suggest that mass loss via winds may be important in sub -Eddington, radiatively inefficient, accretion flows. On the other hand, thermal conduction may play a significant role in such systems (Tanaka \& Menou 2006). Using a simplified model, we included both winds and thermal conduction in unified model in order to understand their possible combined effects on the dynamics of the system. Accounting for a variable mass accretion rate in the flow in proportion to $R^{s}$ and a saturated form of thermal conduction with coefficient $\phi_{\rm s}$, a set of self-similar solutions are presented in this study.  We have varied $s$ and $\phi_{\rm s}$ in our models to judge their sensitivity to these parameters. The most important finding of our analysis is that as more mass, angular momentum and energy flux are taking away from the disc (i.e. stronger winds), possible modifications of the physical profiles due to the thermal conduction will occur at higher level of conduction.

Although our self-similar solutions are too simplified to be used to calculate the emitted spectrum, their typical behavior show importance of thermal conduction in the presence of winds. For future the global solution rather than the self-similar solutions is required. This is because most of the radiation of an ADAF comes from its innermost region where the self-similar solution breaks down.

\acknowledgments
I appreciate the referee, N. Shakura, for his useful comments.
This research  was funded under the Programme for Research in
Third Level Institutions (PRTLI) administered by the Irish Higher
Education Authority under the National Development Plan and with partial
support from the European Regional Development Fund.


\begin{thebibliography}{}

\bibitem{} Abramowicz, M. A., Chen, X., Kato, S., Lasota, J. P., Regev, O.: ApJ, {\bf 438}, L37 (1995)

\bibitem{} Abramowicz, M. A., Czerny, B., Lasota, J. -P., Szuszkiewicz, E.: ApJ, {\bf 332}, 646 (1988)

\bibitem{} Beckert, T.: ApJ, {\bf 539}, 223 (2000)

\bibitem{} Blandford, R. D., Begelman, M. C.: MNRAS, {\bf 303}, L1 (1999)

\bibitem{} Chen, X.: MNRAS, {\bf 275}, 641 (1995)

\bibitem{} Cowie, L. L., McKee, C. F.: ApJ, {\bf 211}, 135 (1977)

\bibitem{} Fukue, J.: PASJ, {\bf 56}, 569 (2004)

\bibitem{} Honma, F.:  PASJ, {\bf 48}, 77 (1996)

\bibitem{} Igumenshchev, I. V., Abramowicz, M. A.,  Narayan, R.:  ApJ, {\bf 537}, L27 (2000)

\bibitem{} Johnson, B. M., Quataert, E.: ApJ, {\bf 660}, 1273 (2007)

\bibitem{} Kitabatake, E., Fukue, J., Matsumoto, K.: PASJ, {\bf 54}, 235 (2002)

\bibitem{} Knigge, C.: MNRAS, {\bf 309}, 409 (1999)

\bibitem{} Lin, D. J., Misra, R., Taam, R. E.: MNRAS, {\bf 324}, 319 (2001)

\bibitem{} Manmoto, T., Kato, S., Nakamura, K. E., Narayan, R.: ApJ, {\bf 529}, 127 (2000)

\bibitem{} Misra, R., Taam, R. E.: ApJ, {\bf 553}, 978 (2001)

\bibitem{} Narayan, R., Igumenshchev, I. V.,  Abramowicz, M. A.:  ApJ, {\bf 539}, 798 (2000)

\bibitem{} Narayan, R., Yi, I.: ApJ, {\bf 428}, L13 (1994)

\bibitem{} Quataert, E., Narayan, R.: ApJ, {\bf 520}, 298 (1999)

\bibitem{} Sharma, P., Hammett, G. W., Quataert, E., Stone, J. M.: ApJ, {\bf 637}, 952 (2006)

\bibitem{}  Shakura,  N. I.,  Sunyaev,  R. A.:  A\&A, {\bf 24}, 337 (1973)

\bibitem{} Tanaka, T., Menou, K.: ApJ, {\bf 649}, 345 (2006)

\bibitem{} Turolla, R., Dullemond, C. P.: ApJ, {\bf 531}, L49 (2000)

\bibitem{} Xie, F. G., Yuan, F.: ApJ, {\bf 681}, 499 (2008)

\bibitem{} Yuan, F., Markoff, S., Falcke, H.: A\&A, {\bf 383}, 854 (2002)





\end{thebibliography}
\end{document}